\documentclass[aps,prd,twocolumn,superscriptaddress,showkeys,showpacs]{revtex4}
\usepackage{pdfsync}
\usepackage{mathptmx}
\usepackage[utf8]{inputenc}
\usepackage[T1]{fontenc}
\usepackage{slashed}
\usepackage{times}
\usepackage{amssymb,amsfonts,amsmath,amsthm}
\usepackage{dsfont,bbm}
\usepackage{dcolumn}
\usepackage{epsf}
\usepackage{graphicx}
\usepackage[caption=false]{subfig}
\usepackage{dsfont}
\usepackage{slashed}
\usepackage[active]{srcltx}
\usepackage[usenames]{color}

\begin{document}

\title{On vacuum integration}

\author{I.~V.~Anikin}
\email{anikin@theor.jinr.ru}
\affiliation{Bogoliubov Laboratory of Theoretical Physics, JINR,
             141980 Dubna, Russia}

\begin{abstract}
The effective potential is known to be given by the vacuum diagrams.
In this paper we show that the massive loop integrations corresponding to the vacuum diagrams
can be expressed through the massless loop integrations with the corresponding massive prefactor
provided the special treatment of $\delta(0)$-singularity. In its turn, the massless loop integrations
provide the useful instrument for the conformal symmetry application.
\end{abstract}
\date{\today}
\maketitle


It is known that the effective potential method is an efficient approach to many subjects, whether it is
the spontaneous symmetry breaking or the critical behaviour theory. From the point of view of Feynman diagrams,
the effective potential is nothing but a sum of the vacuum diagram sets. In its turn, the vacuum diagrams are defined by the
corresponding loop integrations. If the loop integrations are made from the massive (scalar) propagators,
the conformal symmetry is useless and, at first glance, there is no reason to think that a useful instrument
such as conformal symmetry has any benefit.

However, a simple observation, which we describe in the present paper, gives a possibility to
express the massive loop integrations corresponding to the vacuum diagrams
through the massless loop integrations, which are conformal invariant objects, with the corresponding massive prefactor
provided the special treatment of $\delta(0)$-singularity.

For pedagogical reasons, we begin with the illustrative examples that explain the main items of our approach.
Let us first consider the following integration
\begin{eqnarray}
\label{Int-1}
{\cal I}= \int d^D\mu(k) \, \sum\limits_{n=1}^{\infty}\frac{1}{n}\Big[ \frac{m^2}{k^2}\Big]^n
\quad\text{provided}\quad \Big| \frac{m^2}{k^2}\Big| < 1,
\end{eqnarray}
where $m^2$ is an arbitrary dimensionful parameter. 
The integration measure $d^D\mu(k)$ is assumed to be good enough to ensure the full convergency, {\it i.e.}
one can define it, for instance, as
\begin{eqnarray}
\label{def-mi}
\int d^D\mu(k) = \int_{\Omega} d^D k,
\end{eqnarray}
where $\Omega$ implies a certain restricted $D$-dimensional region.

In the following, notice that, for the sake of shortness,
we maximally omit the irrelevant numerical normalization constants together with the corresponding $\mu^{2\epsilon}$
which accompanies any loop integrations in QFT leading to the dimensionless combinations in the final results.

The well-defined ${\cal I}$ can be calculated by two ways: $(a)$
one performs the summation and then calculates the integration over $k$, resulting in ${\cal I}_{(a)}$;
$(b)$ since the integration commutes with the summation as well as the sum is convergent,
one performs the integration over $k$ and then sums the series, resulting in ${\cal I}_{(b)}$.
These two ways merely reflect the theorem on the integration of series and ${\cal I}_{(a)}={\cal I}_{(b)}$.

If $\Omega\to\infty$, the integration may include the divergency and ${\cal I}_{(a)}\not= {\cal I}_{(b)}$.
However, as shown below, despite the finite parts of ${\cal I}_{(a)}$ and  ${\cal I}_{(b)}$ are different in this case,
the singular parts of ${\cal I}_{(a)}$ and  ${\cal I}_{(b)}$ can be equal.
Anticipating the QFT applications of the proposed method,  we deal with only the singular parts
in the present paper.

Let us first consider ${\cal I}_{(a)}$.
The first way $(a)$ is rather standard and deals with the loop integration involving the massive (scalar) propagators. Making use of
($|x|<1$ for the convergency)
\begin{eqnarray}
\label{Ln-f}
- \ln(1-x) = \sum\limits_{n=1}^{\infty}\frac{x^n}{n},
\end{eqnarray}
we readily derive that
\begin{eqnarray}
\label{Int-2}
&&{\cal I}_{(a)}=- \lim\limits_{\Omega\to\infty}\int_\Omega d^D\mu(k) \, \ln\frac{(k^2-m^2)/\mu^2}{k^2/\mu^2}=
\nonumber\\
&&
i \frac{\partial}{\partial \alpha}\Big\{ \int d^D k_E
\Big[ \frac{\mu^{2\alpha}}{\left(k^2_E+m^2\right)^\alpha} - 
\frac{\mu^{2\alpha}}{\left(k^2_E\right)^\alpha} \Big]  \Big\} \Big|_{\alpha=0}
\nonumber\\
&&=
i \pi^{D/2} \Gamma(-D/2) (m^2)^{D/2}.
\end{eqnarray}
Notice that in Eqn.~(\ref{Int-2}) one deals with the massive loop integration.
Also, one can see that if $D/2\in \mathbb{Z}$ the considered integration has the singularity
corresponding first-order poles due to the presence of $\Gamma$-function.

We are now in position to discuss ${\cal I}_{(b)}$. This way is based on the nontrivial representation of the massless vacuum integration \cite{Gorishnii:1984te}.
We have
\begin{eqnarray}
\label{Int-3}
{\cal E}=\int d^D k \, \frac{1}{\big(k^2\big)^n}=
\frac{2i\, \pi^{1+D/2}}{(-1)^n\, \Gamma(n)} \delta(n-D/2).
\end{eqnarray}
Notice that Eqn.~(\ref{Int-3}) can also be written in terms of Fourier transforms as
\begin{eqnarray}
\label{Int-3-2}
&&{\cal E}=\int (d^D k) \Big\{ C^{-1}(D,n) \int d^Dz\, e^{-i k z}\frac{1}{\big(z^2\big)^{D/2-n}} \Big\}
\nonumber\\
&&=C^{-1}(D,n) \int d^Dz \,\delta(z) \,\frac{1}{\big(z^2\big)^{D/2-n}},
\end{eqnarray}
where
\begin{eqnarray}
\label{C}
C^{-1}(D,n)=i(-\pi)^{D/2}\frac{\Gamma(D/2-n)}{\Gamma(n)}.
\end{eqnarray}
If $D/2-n=0$ then the {\it r.h.s.} of Eqn.~(\ref{Int-3-2}) becomes a dimensionless one and as a consequence ${\cal E}\not= 0$.
At the same time, if  $D/2-n\not= 0$, the dimensional analysis results in ${\cal E}=0$ as it must be.

Hence, taking into account Eqn.~(\ref{Int-3}),  ${\cal I}_{(b)}$ has a form of
\begin{eqnarray}
\label{Int-4}
&&{\cal I}_{(b)}=\lim\limits_{\Omega\to\infty}
\sum\limits_{n=1}^{\infty}\frac{(m^2)^n}{n} \int_\Omega d^D\mu(k) \,\frac{1}{(k^2)^n}
\nonumber\\
&&=\frac{2i \pi^{D/2+1}}{\Gamma(D/2)} \sum\limits_{n=1}^{\infty} \frac{(-1)^{-n} (m^2)^n}{n}\, \delta(n-D/2).
\end{eqnarray}

One can see that if $\Omega\to\infty$ then ${\cal I}$ has two
equivalent representations given by Eqns.~(\ref{Int-2}) and (\ref{Int-4}), {\it i.e.}
\begin{eqnarray}
\label{Equiv-1}
&&\Gamma(-D/2) (m^2)^{D/2}=
\nonumber\\
&&\frac{2\pi}{\Gamma(D/2)} \sum\limits_{n=1}^{\infty} \frac{(-1)^{-n} (m^2)^n}{n}\, \delta(n-D/2)
\end{eqnarray}
provided $n=D/2$.
In Eqn.~(\ref{Equiv-1}), the argument of $\delta$-function
singles out $n=D/2$ giving the only nonzero term of the sum over $n$ which involves $\delta(0)$.

We emphasize that in contrast to the way $(a)$, the second way $(b)$ resulting in Eqn.~(\ref{Int-4}) deals with
the conformal invariant massless integrations over $k$.

We now dwell on the relation given by Eqn.~(\ref{Equiv-1}) which is valid for the integer value of $D/2$.
Eqn.~(\ref{Equiv-1}) can simply be presented as
\begin{eqnarray}
\label{G-rep}
\Gamma(- D/2) = \frac{2\pi (-1)^{-D/2}}{\Gamma(D/2+1)} \, \delta(0) \quad \text{with}\quad D/2\in \mathbb{Z}
\end{eqnarray}
which hints that the singularity corresponding to the pole of $\Gamma$-function can be traded for the singularity like $\delta(0)$.

Indeed, as well-known from \cite{G-Sh}, the $\delta$-function as a singular distribution (generalized) function
can be treated as a certain limit of the delta-like function sequences,
\begin{eqnarray}
\label{fun-delta}
f_\nu(t)=\frac{1}{\pi} \frac{\nu}{t^2+\nu^2} \quad (\nu>0).
\end{eqnarray}
That is, we have
\begin{eqnarray}
\label{d-2}
\frac{1}{\pi}\lim\limits_{\nu\to 0}\frac{\nu}{t^2+\nu^2}\equiv
\frac{1}{\pi}\lim\limits_{\nu\to 0} \Im\text{m}\frac{1}{t-i\nu}=\delta(t).
\end{eqnarray}
At the same time, the $\delta$-function has the following representation in terms of the Fourier transform,
\begin{eqnarray}
\label{d-2-2}
2\pi \delta(t)=\int\limits_{-\infty}^{+\infty} d z\, e^{-i z t}.
\end{eqnarray}
Therefore, taking into account Eqns.~(\ref{d-2}) and (\ref{d-2-2}), we can define the singularity of $\delta(0)$-type as
\begin{eqnarray}
\label{d-3}
\pi \delta(0)=\frac{1}{2}\int_{-\infty}^{+\infty} d z = \lim\limits_{\nu\to 0}\frac{1}{\nu}=\infty.
\end{eqnarray}
Now if we consider the case of $D=4-2\epsilon$ where $\epsilon\to 0$, Eqn.~(\ref{Equiv-1}) takes the form
\begin{eqnarray}
\label{Con-1}
\lim\limits_{\epsilon\to 0} \frac{\Gamma(\epsilon)}{(2-\epsilon)(1-\epsilon)} m^{-2\epsilon} =
2\pi \delta(0) \lim\limits_{\epsilon\to 0} \frac{m^{-2\epsilon}}{\Gamma(3-\epsilon)}.
\end{eqnarray}
In other words, with the help of Eqn.~(\ref{d-3}), we can infer that
\begin{eqnarray}
\label{dd-2}
\lim\limits_{\epsilon\to 0} \Gamma(\epsilon) = 2\pi \delta(0) = \lim\limits_{\epsilon\to 0} \frac{1}{\epsilon}.
\end{eqnarray}
and, multiplying by $\mu^{2\epsilon}$, we can finally write that
\begin{eqnarray}
\label{Con-2}
2\pi \lim\limits_{\epsilon\to 0}\delta(\epsilon) \Big[\frac{m^2}{\mu^2}\Big]^{-\epsilon}
\Rightarrow \Big[ \frac{1}{\epsilon} - \ln\frac{m^2}{\mu^2}\Big]
\Leftarrow \lim\limits_{\epsilon\to 0} \Gamma(\epsilon) \Big[\frac{m^2}{\mu^2}\Big]^{-\epsilon}.
\end{eqnarray}

It is instructive to present the alternative treatment of $\delta(0)$-singularity in Eqn.~(\ref{dd-2}).
The condition $|m^2/k^2|<1$ of Eqn.~(\ref{Int-1}) ensures that the infrared divergency, $|k^2|\to 0$, is absent.
Therefore, focusing on the ultraviolet divergency, $|k^2|\to \infty$, and excluding the special case of $n=D/2$, see Eqn.~(\ref{Int-3}),
we get the following \cite{Grozin:2007zz} (here, $D=4-2\epsilon$)
\begin{eqnarray}
\label{uv-div}
\int_{\overline{\Omega}_{UV}}  \frac{d^D k}{(-k^2)^n}\Big|_{n\not= D/2} =\frac{i (2\pi)^{D-2}}{4} \frac{1}{\epsilon}
\end{eqnarray}
which also supports Eqn.~(\ref{dd-2}).

Thus, the first way $(a)$ and the second way $(b)$ give the same final result iff the $\delta(0)$-singularity
has been treated as in Eqn.~(\ref{dd-2}). This is our principal observation.

%
%
\begin{figure}[t]
\centerline{\includegraphics[width=0.2\textwidth]{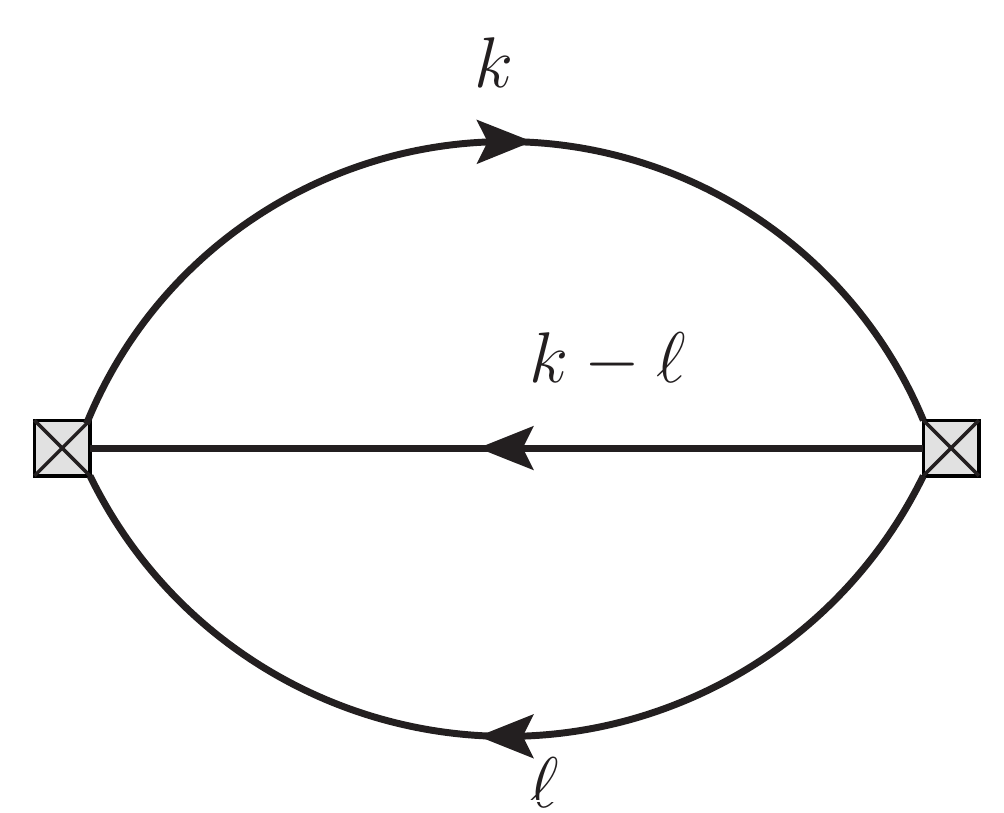}}
\caption{The sunset-type diagram.}
\label{Fig-1}
\vspace{-0.5cm}
\end{figure}
As a simplest example of how the method based on Eqn.~(\ref{Int-3}) works in practice, let us consider
the so-called vacuum sunset diagram, see Fig.~\ref{Fig-1}, which is given by
\begin{eqnarray}
\label{v-s-1}
&&{\cal J}= \int \frac{d^D k}{k^2} \, \int d^D \ell\frac{1}{\ell^2(k-\ell)^2}\Rightarrow
\\
&&G(1,1) \int \frac{d^D k}{(k^2)^{3-D/2}} \Rightarrow G(1,1) \delta(3-D),
\nonumber
\end{eqnarray}
where \cite{Grozin:2007zz, Chetyrkin:1980pr}
\begin{eqnarray}
\label{G-1}
G(1,1)=\frac{\Gamma(2-D/2) \Gamma^2(D/2-1)}{\Gamma(D-2)} \sim \frac{1}{\epsilon}.
\end{eqnarray}
In Eqn.~(\ref{G-1}), the symbol $\sim$ means ``behaves like''
\footnote{We say that the function
$f(x)$ behaves like the function $g(x)$ as $x\to a$, {\it i.e.} $f(x)\sim g(x)$, if
$\lim\limits_{x\to a} f(x)/g(x)=1$.}.
Hence, we get that
\begin{eqnarray}
\label{v-s-2}
{\cal J}\sim \frac{\delta(1)}{\epsilon} = 0,
\end{eqnarray}
which is in agreement with the usual expectation.

\begin{figure}[t]
\centerline{\includegraphics[width=0.2\textwidth]{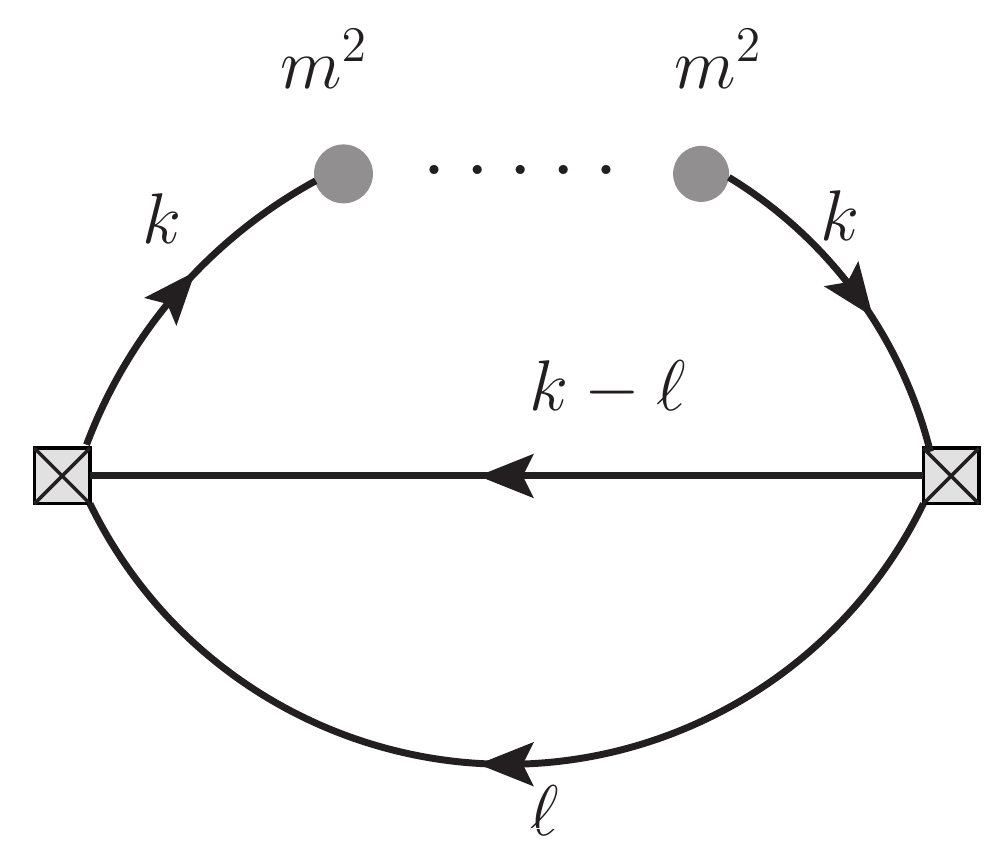}}
\caption{The sunset-type diagram with $(m^2)^n$ insertion in the upper scalar propagator of the diagram depicted in Fig.~\ref{Fig-1}.}
\label{Fig-2}
\end{figure}

The next illustrative example is given by the following integration, see the diagram depicted in Fig.~\ref{Fig-2},
\begin{eqnarray}
\label{K1-int-1}
{\cal K}_1=\sum\limits_{n=1}^{\infty}  \int d^D k\frac{(m^2)^n}{(k^2)^{n+1}} \, \int d^D \ell\frac{1}{\ell^2(k-\ell)^2}.
\end{eqnarray}
This integration corresponds to the sum of vacuum sunset-like diagrams if the set $(m^2)^n$ has been inserted
into the upper scalar line (the massless propagator) of the original sunset diagram.
Having calculated the integration over $d^D\ell$ in the integrand, we write that
\begin{eqnarray}
\label{Int-ell}
\int d^D \ell\frac{1}{\ell^2(k-\ell)^2} \Rightarrow (k^2)^{D/2-2} G(1,1).
\end{eqnarray}
Notice that in Eqn.~(\ref{Int-ell})  the divergency generated by $G(1,1)\sim 1/\epsilon$
 has to be subtracted as a sub-divergency of the diagram.

\begin{widetext}
Therefore, the integral ${\cal K}_1$ reads
\begin{eqnarray}
\label{K1-int-2}
&&{\cal K}_1= \sum\limits_{n=1}^{\infty}  \int d^D k\frac{G(1,1)\,(m^2)^n}{(k^2)^{n+3-D/2}}
-\frac{2}{\epsilon}\sum\limits_{n=1}^{\infty}  \int d^D k\frac{(m^2)^n}{(k^2)^{n+1}}
\Rightarrow
 \sum\limits_{n=1}^{\infty} \Big[\frac{m^2}{\mu^2}\Big]^n \Big[ \delta(n+3-D) G(1,1)
- \frac{2}{\epsilon} \delta(n+1-D/2) \Big]
\nonumber\\
&&\Rightarrow \frac{1}{\epsilon}\Big[\frac{m^2}{\mu^2}\Big]^{D-3} \delta(0)-
\frac{2}{\epsilon}\Big[\frac{m^2}{\mu^2}\Big]^{D/2-1} \delta(0)
\Rightarrow - \frac{m^2}{\mu^2} \Big[\frac{1}{\epsilon^2} - \ln^2 \frac{m^2}{\mu^2} \Big].
\end{eqnarray}
\end{widetext}

\begin{figure}[t]
\centerline{\includegraphics[width=0.2\textwidth]{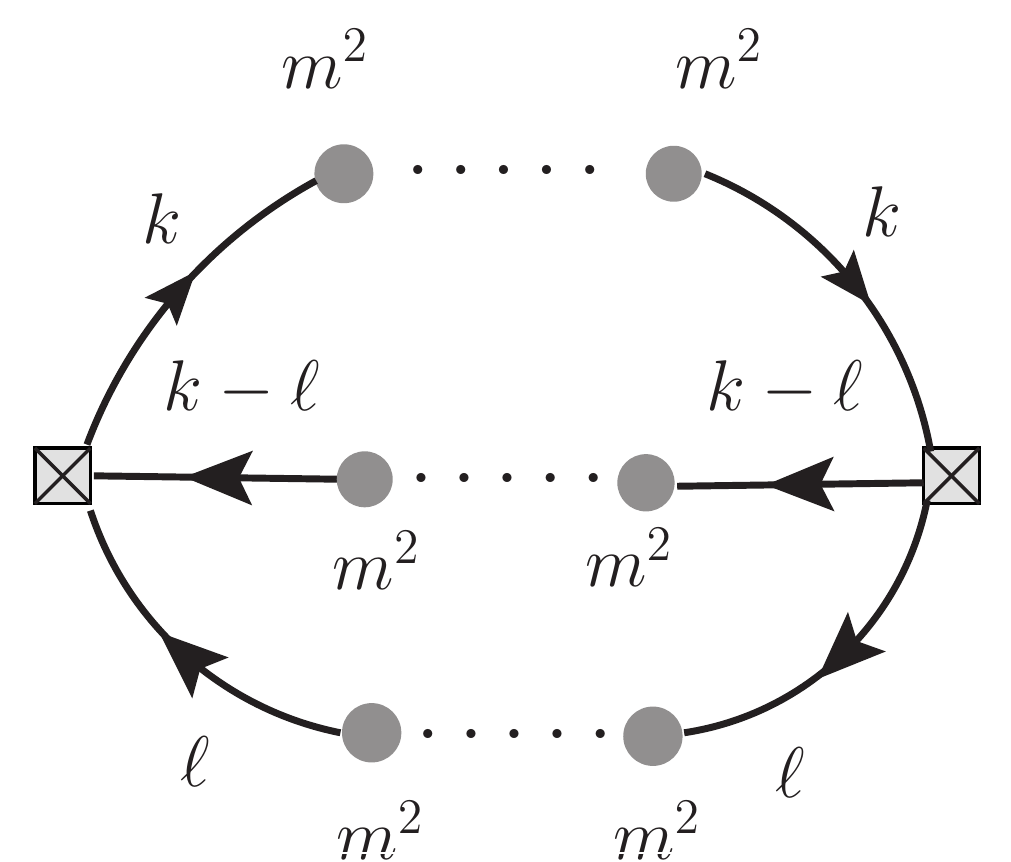}}
\caption{The sunset-type diagram with $(m^2)^n$ insertion in each scalar propagator of the diagram depicted in Fig.~\ref{Fig-1}.}
\label{Fig-3}
\end{figure}

And, the last example is associated with the following integration
\begin{eqnarray}
\label{K2-int-1}
&&{\cal K}_2=\sum\limits_{n_1+n_2+n_3=1}^{\infty}  \int d^D k \frac{(m^2)^{n_1+n_2+n_3}}{(k^2)^{n_1+1}} \,
\nonumber\\
&&\times
\int d^D \ell\frac{1}{\left(\ell^2\right)^{n_2+1}\left((k-\ell)^2\right)^{n_3+1}}.
\end{eqnarray}
This integration generalizes the third example, and it corresponds to the diagrams where the $(m^2)^{n_i}$-insertion has been
implemented into each scalar line of the sunset-like diagram, see Fig.~\ref{Fig-3}.
Again, we first calculate the integration over $d^D\ell$ in the integrand resulting in
\begin{eqnarray}
\label{Int-ell-2}
&&\int d^D \ell\frac{1}{\left(\ell^2\right)^{n_2+1}\left((k-\ell)^2\right)^{n_3+1}}
\nonumber\\
&&
\Rightarrow (k^2)^{D/2-n_2-n_3-2} G(n_2+1,n_3+1).
\end{eqnarray}
As in Eqn.~(\ref{Int-ell}), in Eqn.~(\ref{Int-ell-2}) the divergency generated by (in what follows $n=n_1+n_2+n_3$)
\begin{eqnarray}
\label{G-exp-4}
\sum\limits_{n=1}^{\infty}G(n_2+1,n_3+1)\Big|_{n=D-3}\sim 1/\epsilon
\end{eqnarray}
is to be subtracted as a sub-divergency of the diagram.

\begin{widetext}
Hence, the integral ${\cal K}_2$ takes the form
\begin{eqnarray}
\label{K2-int-2}
&&{\cal K}_2= \sum\limits_{n=1}^{\infty}
\int d^D k\frac{G(n_2+1,n_3+1)\, (m^2)^{n}}{(k^2)^{n+3-D/2}}
- \frac{2}{\epsilon}\sum\limits_{n=1}^{\infty}
\int d^D k\frac{(m^2)^{n}}{(k^2)^{n_1+1}}
\nonumber\\
&&
\Rightarrow
 \sum\limits_{n=1}^{\infty}\Big[\frac{m^2}{\mu^2}\Big]^n \delta(n+3-D) \,G(n_2+1,n_3+1)
-\frac{2}{\epsilon}\sum\limits_{n=1}^{\infty} \Big[\frac{m^2}{\mu^2}\Big]^n \delta(n_1+1-D/2)
\Rightarrow - \frac{m^2}{\mu^2} \Big[\frac{1}{\epsilon^2} - \ln^2 \frac{m^2}{\mu^2} \Big].
\end{eqnarray}
\end{widetext}

We stress that the considered vacuum integrations ${\cal I}$, ${\cal K}_1$ and ${\cal K}_2$ appear as typical integrations when the
effective potential and its evolution in $\lambda\varphi^4$-theory have been calculated within the generating functional approach
where the mass (or the other dimensionful parameter) terms of Lagrangian
are assumed to be as ``interaction'' terms forming the effective interaction vertices
provided all (scalar) propagators are massless \cite{Anikin-prog}.
In particular, calculating the generating functional in $\lambda\varphi^4$-theory by the 
stationary phase method, we obtain the following asymptotical expansion 
\begin{eqnarray}
\label{Z-st-ph}
&&{\mathbb Z}[J]= e^{iS(\varphi_c) + i (J,\,\varphi_c)} \int ({\cal D}\varphi)
 e^{(\varphi,\, \Box\varphi)}
\nonumber\\
&&
\times
\Big\{1 +
m^2_e(\varphi_c) \int dz \varphi^2(z) +
\lambda\varphi_c \int dz \varphi^3(z) +
\lambda\int dz \varphi^4(z) 
\nonumber\\
&&
+ \big(\text{terms with} \,\,\, [m^2_e(\varphi_c)]^n,  [\lambda\varphi_c ]^n, [\lambda]^n \,\,\,  | \,\,\,  n>1\big) \Big\}
\end{eqnarray}
where $m^2_e(\varphi_c)=m^2+\lambda\varphi_c^2/2$ with $\varphi_c$ being the classical field.
Hence, one generates the Feynman rules with
three types of interaction vertices 
\begin{eqnarray}
\label{ver-1}
V_2 = m^2_e(\varphi_c) \varphi^2;\quad
V_3 = \lambda\varphi_c \varphi^3;\quad
V_4 = \lambda\varphi^4
\end{eqnarray}
and all inner lines correspond to the scalar {\it massless} propagators.

With the help of Eqn.~(\ref{Z-st-ph}),
having replaced $m^2\to m^2_e(\varphi_c)$
in Eqn.~(\ref{Con-2}) and having restored the numerical coefficients,
we get the effective potential ${\cal V}[\varphi_c]$ in the form of
(to the one-loop accuracy)
\begin{eqnarray}
\label{Eff-Pot-ee-1}
&&
{\cal V}[\varphi_c, \mu^2|\, {\cal I}]={\cal L}(\varphi_c) +
\frac{m^4_e(\varphi_c)}{4(4\pi)^2} \ln\frac{m^2_e(\varphi_c)}{\mu^2}
\nonumber\\
&&
+
(\ell\text{-loop contr.}\,\,\, |\,\,\, \ell>1) ,
\end{eqnarray}
where
${\cal L}(\varphi_c)$ is the mass and interaction parts of Lagrangian in $\lambda \varphi^4$-theory.

In conclusion, we outline the main result of the present paper as the following.
At the beginning, we deal with some finite and restricted integration region $\Omega$ in Eqn.~(\ref{Int-1}).
The conditions of theorem on integration of the sum are fulfilled resulting
in ${\cal I}_{(a)} = {\cal I}_{(b)}$.
Then, if $\Omega \to \infty$, ${\cal I}_{(a)}$ and ${\cal I}_{(b)}$
are not equal and, moreover, they become singular.
It turns out that both ${\cal I}_{(a)}$ and ${\cal I}_{(b)}$ have the same type 
of singularities, see Eqn.~(\ref{G-rep}).
Indeed, $\delta(0)$-singularity is nothing but $1/\epsilon$ at $\epsilon \to 0$.
On the other hand, $\Gamma$-function has only first-order poles given by the same $1/\epsilon$.
So, instead of the massive loop integrations, we can deal with the
massless loop integrations. Both cases can be renormalized in the same way.

Thus, we have demonstrated the method
which allow us to reduce the calculation of massive loop integrations corresponding to the vacuum diagrams
to the calculation of the massless (conformally invariant) loop integrations which is more simple, from the technical point of view.
To this goal, we have proposed the special treatment of $\delta(0)$-singularity which can be related to the
pole of $\Gamma$-function.

We thank A.B.~Arbuzov, M.~Deka, A.L.~Kataev, S.V.~Mikh\-ailov, L.~Szymanowski, A.S.~Zhevlakov  
and the colleagues from the Theoretical Physics Division of NCBJ (Warsaw)
for useful discussions.


\begin{thebibliography}{99}
\vspace{1\baselineskip}

\bibitem{Gorishnii:1984te}
  S.~G.~Gorishnii and A.~P.~Isaev,
  Theor.\ Math.\ Phys.\  {\bf 62}, 232 (1985)
  [Teor.\ Mat.\ Fiz.\  {\bf 62}, 345 (1985)].

\bibitem{G-Sh}
I.~M.~Gel'fand and G.~E.~Shilov,
``Generalized Functions'', V. 1, 1964

\bibitem{Grozin:2007zz}
  A.~Grozin,
  ``Lectures on QED and QCD: Practical calculation and renormalization of one- and multi-loop Feynman diagrams,''
  Hackensack, USA: World Scientific (2007) 224 p

\bibitem{Chetyrkin:1980pr} 
  K.~G.~Chetyrkin, A.~L.~Kataev and F.~V.~Tkachov,
  Nucl.\ Phys.\ B {\bf 174}, 345 (1980).

\bibitem{Anikin-prog}
I.~V.~Anikin, work in progress.

\end{thebibliography}
\end{document}